\documentclass[11pt]{article}
\usepackage{geometry}                
\usepackage{graphicx}
\title{Homage to Ettore Majorana\footnote{Majorana Memorial, Catalina, Italy, October 2006}}
\author{R. Jackiw\\
\it \small Department of Physics\\
\it \small Massachusetts Institute of Technology\\
\it \small Cambridge, MA 02139\\
\small\tt MIT-CTP-3779
}
\date{}                                           
\begin{document}
\maketitle
\thispagestyle{empty}
I am privileged to be in this interesting place honoring Ettore Majorana. Of course I have had no personal contact with him --- he disappeared before I appeared. However,  it is surprising, that I  did not encounter his name nor his achievements during my physics education. He is hardly mentioned in the usual textbooks, at least in the American ones. This is a great loss! To me it was a special loss for the following reason. When I was preparing with Hans Bethe our quantum mechanics textbook, I became fascinated by the Thomas-Fermi theory, and I strived to give a complete discussion in our text. But at that time I knew nothing about Majorana's work in this area, and so could not include it. Again, when we were writing the chapters on Dirac theory, I wondered why only charged fermions are considered. The resolution of my puzzlement lay in the Majorana representation, about which I learned only later, principally through Julian Schwinger's writings. Schwinger apparently appreciated the Majorana approach and in his discussions of Dirac theory, the charge carrying fermion field is usually presented as the complex superposition of two  real fields, in complete analogy to the description of charged boson fields. Another connection between Majorana and Schwinger can be noted. The last topic that Schwinger researched concerned corrections to the Thomas-Fermi model.

Schwinger was also interested in the problem of mass generation, a topic which these days is linked to Majorana's name. I shall use this point of contact between the two scientists to review Schwinger's mass generation mechanism \cite{Schwinger:1962tp}, to expose its topological underpinings and to present an interesting generalization \cite{Dvali:2006ps}. Majorana deconstructed the complex Dirac equation into its real components. Here I deconstruct Schwinger's mass generation into its topological ingredients. I think that Majorana would have liked these results.

\section{ Schwinger Model Resum\'{e}}

     In the Schwinger model, an Abelian vector potential    $A_{\mu}$      interacts with a vector current         $\mathcal{J}^{\mu}$ constructed from massless Dirac fields $\psi$. The Lagrange density is gauge invariant and reads
\begin{equation}
\label{lagrangian}
\mathcal{L} \, = \, -{1\over 4} F^{\mu\nu}F_{\mu\nu}\, + \, i \bar{\psi} \gamma_{\mu} (\partial_{\mu}
\, + \, i e  A_{\mu})\psi, ~~~~ F_{\mu\nu} \equiv  \partial_{\mu} A_{\nu} \, - \, \partial_{\nu}A_{\mu},
\end{equation}
\begin{equation}
\label{l1}
\mathcal{L}_I \, = \, - e \mathcal{J}^{\mu} A_{\mu}, ~~~ \mathcal{J}^{\mu} \, = \, \bar{\psi}\gamma^{\mu}\psi.
\end{equation}
The model is defined on an unphysical 2-dimensional space-time, where the Dirac matrices are $2+2$ and the Dirac spinor $\psi$ possesses tow components. The traditional solution of the model proceeds by functionally integrating the Dirac fields, giving an effective action.
\begin{equation}
\label{ieff}
\mathcal{I}_{eff}(A) \, = \, -{1\over 4} \int F^{\mu\nu} F_{\mu\nu} \, - \, i \
ln \ det [\gamma^{\mu}(\partial_{\mu}\, + \, i e A_{\mu})]
\end{equation}
The functional determinant can be computed because the only non vanishing Feynman diagram is the vacuum polarization graph. (This is a special feature of two dimensions.)  \\
\centerline{\includegraphics[scale=.85]{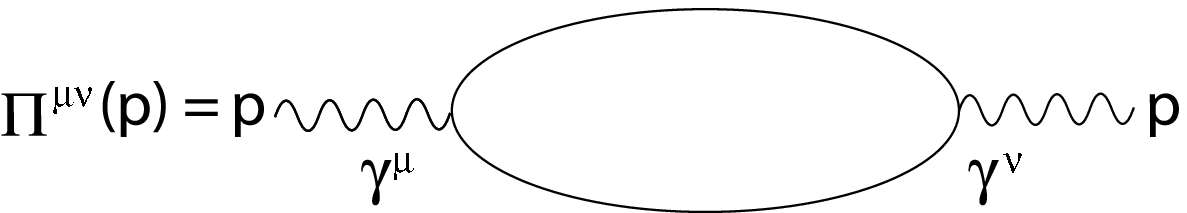}}\label{PDFpar}
{\small Figure 1: Vacuum polarization graph generates the polarization operator
$\Pi^{\mu\nu}(p) \, \propto \, (g^{\mu\nu} \, - \, p^{\mu}p^{\nu}/p^2)$.}\\[1.5ex]
This generates the polarization tensor 
$\Pi^{\, \mu\nu}(p) \, \propto \, (g^{\mu\nu} \, - \, p^{\mu}p^{\nu}/p^2)$. 
The coefficient of  $g^{\mu\nu}$  is evaluation dependent (the diagram is superficially divergent),
but it becomes fixed by the gauge invariance requirement that the vector current correlator (whose proper part is   $\Pi^{\ \mu\nu}$) be transverse. The effective action 
\begin{equation}
\label{ieff1}
\mathcal{I}_{eff}(A) \, = \, \int \left [ -\, {1 \over 4}  F^{\mu\nu} F_{\mu\nu} \, 
+  \, {e^2 \over 2\pi}  A_{\mu}(g^{\mu\nu}  \, - \, {\partial^{\mu}\partial^{\nu} \over \partial^2} ) A_{\nu}\right ],
\end{equation}
exhibits the generated mass, $m^{\, 2} \, = \, {e^2 \over \pi}$. Thus Schwinger showed that a gauge invariant theory may nevertheless possess a mass gap --- a result known to superconductivity experts, as emphasized by Philip Anderson.   Although usually one says that the ``photon'' acquires a mass, in two dimensions the ``photon'' field  $A_{\mu}$  can be decomposed as                       
 $A_{\mu} \, = \, \partial_{\mu}\theta  \, + \, \epsilon_{\mu\nu}\partial^{\nu}\eta'$.                                                The gauge part decouples; only the pseudoscalar  $\eta'$  remains. So one could just as well say that a pseudoscalar excitation acquires the mass.
     
     It is important to appreciate that the axial vector current   $\mathcal{J}_{\alpha}^{\, 5} = \bar{\psi}\,  \gamma_\alpha \gamma^5 \psi$,  which is conserved with massless fermions with unquantized $\psi$, acquires an anomalous divergence upon quantization. This is immediately seen when the 2-dimensional duality relation between  axial and vector currents is used.
\begin{equation}
\label{axialvector}
\mathcal{J}_{\alpha}^5\, = \, \epsilon_{\alpha\mu} \mathcal{J}^{\mu}
\end{equation}
Formula (\ref{axialvector}) is a consequence of 2-dimensional geometry: when   $\mathcal{J}^{\mu} $   is a vector,   $\mathcal{J}_{\alpha}^5$   defined by (\ref{axialvector}) is an axial vector. More explicitly, 
(\ref{axialvector}) is seen in a 2-dimensional gamma matrix identity.
\begin{equation}
\label{gamma}
\gamma_{\alpha}\gamma^5 \, = \, \epsilon_{\alpha\mu}\gamma^{\mu}
\end{equation}
Therefore, the correlator ${^5\Pi}_{\alpha}^{\ \nu}$ of $\mathcal{J}_{\alpha}^5$  with   $\mathcal{J}^{\nu} $ can be simply 
obtained from $\Pi^{\mu \nu}$  as $^5\Pi_{\alpha}^{\ \nu} \, = \, \epsilon_{\alpha\mu} \Pi^{\mu\nu}$.
Moreover, once a transverse form for  $\Pi^{\mu\nu}$  is fixed by gauge invariance,  
$^5\Pi_{\alpha}^{\ \nu}$ fails to be transverse in the $\alpha$  index; the divergence of the axial vector current is anomalous.
\begin{equation}
\label{anomaly}
\partial^{\alpha} \mathcal{J}_{\alpha}^5 \, = \, -{e \over 2\pi} \epsilon^{\mu\nu} F_{\mu\nu}\, = \, 
{e \over \pi} F
\end{equation}
In the second equality we have introduced the (pseudo) scalar  $F$,  dual in two dimensions to the anti symmetric   $F_{\mu\nu} \, \equiv \, \epsilon_{\mu\nu} F$    

     The anomaly provides an immediate derivation of the mass \cite{Farhi:ws1982}. We begin with the gauge field equation of motion that follows from (\ref{lagrangian}).
\begin{equation}
\label{equation}
\partial_{\mu} F^{\mu\nu} \, = \, e \mathcal{J}^{\nu}
\end{equation}
In terms of the dual field strength  $F$   this reads
\begin{equation}
\label{equation1}
\epsilon^{\mu\nu} \partial_{\mu} F \, = \, e \mathcal{J}^{\nu}.
\end{equation}
The   $\epsilon$   symbol may be transferred to the right side  and  $\mathcal{J}^{\nu}$     becomes replaced by its dual $\mathcal{J}_{\alpha}^5$.
  \begin{equation}
  \label{equationdual}
 \partial_{\alpha} F \, = \, -\, e \mathcal{J}_{\alpha}^{5}
\end{equation}
A further divergence gives the d'Alembertian on the left and the anomaly (\ref{anomaly}) on the right.
\begin{equation}
\label{Fequation}
\partial^2 F \, + \, {e^2 \over \pi} F \, = \, 0
\end{equation}
This demonstrates that the pseudoscalar  $F$  acquires a mass, $m^{\, 2} \, = \, {e^2 \over \pi}$.

\section {Topological Entities in the Schwinger Model}

     The 2-dimensional anomaly is proportional to  $- F \, = \, {1\over 2} \epsilon^{\mu\nu}
 F_{\mu\nu}$,  which is recognized as the 2-dimensional Chern-Pontryagin density  
 $\mathcal{P}_{\, 2}$.        
\begin{equation}
\label{p2}
\mathcal{P}_2 \, = \, {1 \over 2} \, \epsilon^{\mu\nu} F_{\mu\nu}.
\end{equation}
Furthermore, the gauge potential  $A_{\mu}$  is dual to the Chern-Simons current      
$\mathcal{C}_{\ 2}^{\alpha}$,  
\begin{equation}
\label{c2duality }
\mathcal{C}_2^{\alpha} \, \equiv \, \epsilon^{\alpha\mu} A_{\mu},
\end{equation}
whose divergence forms the Chern-Pontryagin density \cite{'tHooft:2005cq}.
\begin{equation}
\label{cdiv}
\partial_{\alpha} \mathcal{C}_2^{\alpha} \, = \,\epsilon^{\alpha\mu} \partial_{\alpha} A_{\mu} \, = 
\, {1\over 2} \epsilon^{\alpha\mu} F_{\alpha\mu} \, = \, \mathcal{P}_2
\end{equation}

     The bosonic portion of the Lagrange density for the Schwinger model may be written in terms of these topological entities.

\begin{eqnarray}
\label{p2lagrangian}
\mathcal{L}_2 \, &= &\, -{1\over 4} F^{\mu\nu}F_{\mu\nu}\,  - e \, \mathcal{J}^{\mu} A_{\mu} \, = 
{1\over 2} F^2 \, - \, e A_{\mu} \epsilon^{\mu\alpha} \mathcal{J}_{\alpha}^5 \\ \nonumber 
&=& \, 
{1\over 2} \, \mathcal{P}_2^2 \, + \, e \, \mathcal{C}_2^{\alpha} \mathcal{J}_{\alpha}^5
\end{eqnarray}
Moreover, since  $\mathcal{C}_2^{\alpha}$  and $A_{\mu}$   are linearly related, it makes no difference which one is the fundamental variable. Thus varying   $\mathcal{C}_2^{\alpha}$   in 
(\ref{p2lagrangian}) gives (\ref{equationdual}) directly as the equation of motion.
\begin{equation}
\label{p2equation}
-\partial_{\alpha} \mathcal{P}_2 \, + \, e\, \mathcal{J}_{\alpha}^5 \, = \, 0
\end{equation}
A further divergence and the anomaly equation (\ref{anomaly}) reproduce (\ref{Fequation}), since 
$\mathcal{P}_{\, 2} \, =  \, - \,  F$.    

 It is this last, topological reformulation of the Schwinger model that we shall take to four dimensions. However, we must still address an important point that will arise in the 4-dimensional theory.
     Observe that the equation of motion (\ref{equationdual}) or (\ref{p2equation}) entails an integrability condition: Since the (axial) vector  $\mathcal{J}_{\alpha}^5$ is set equal to a gradient of (the pseudoscalar) $\mathcal{P}_2$,
it must be that the curl of the axial vector vanishes. Equivalently, the dual of the axial vector must be divergence-free; {\it viz.} the vector current must be conserved. Of course the same integrability condition is seen in the original vector formulation of the model, with equation of motion (\ref{equation}), which entails conservation of the vector current  (dual to the axial vector current).
   
     But let us suppose that we have dynamical information only about the topological variables, and do not know whether the current dual to the axial vector current is conserved. (This is the situation that we shall meet in four dimensions.) Then we must reformulate our theory in such a way that the integrability condition is avoided.
     
     This reformulation in two dimensions proceeds by introducing two St\"uckelberg fields $\textsl{p}$
       and $\textsl{q}$   into   $\mathcal{L}_2$.
 \begin{equation}
\label{stuck2}
\mathcal{L}_2' \, = \,  {1 \over 2} \mathcal{P}_2^2 \,  + \, e (\mathcal{C}_2^{\alpha} \, + \,
\epsilon^{\alpha\beta} \partial_{\beta} \textsl{p}) (\mathcal{J}_{\alpha}^5 \, + \, \epsilon_{\alpha\gamma}
\partial^{\gamma} \textsl{q})
\end{equation}    
Upon varying $\mathcal{C}_2^{\alpha}$,  (\ref{p2equation}) becomes replaced by
\begin{equation}
\label{p2equation1}
-\partial_{\alpha} \mathcal{P}_2 \, + \, e\, (\mathcal{J}_{\alpha}^5 \, + \, \epsilon_{\alpha\gamma} \partial^{\gamma} \textsl{q}) \, = \, 0.
\end{equation}
Additionally, variation of  \textsl{p}     and  \textsl{q}    give, respectively
\begin{equation}
\label{peq}
\partial_{\alpha} \epsilon^{\alpha\beta} \mathcal{J}_{\beta}^5 \, + \, \partial^2 \textsl{q}\, = \, 0,
\end{equation}
\begin{equation}
\label{qeq}
\partial^{\alpha} \epsilon_{\alpha\beta} \mathcal{C}_2^{\beta} \, + \, \partial^2 \textsl{p}\, = \, 0.
\end{equation}

The integrability condition on (\ref{p2equation1}) demands that the curl of 
$\mathcal{J}_{\, \alpha}^{\, 5} \, + \, \epsilon_{\alpha\gamma}\partial^{\gamma} \textsl{q}$             
vanish, but this is secured by (\ref{peq}). This equation determines a non-trivial value for  
$\textsl{q}$    if the curl of $\mathcal{J}_{\, \alpha}^{\, 5}$  is non-vanishing, while (\ref{qeq}) fixes an innocuous value for  $\textsl{p}$.   Finally we observe that the divergence of (\ref{p2equation1}) annihilates the  $\textsl{q}$  - dependent term, leaving in the end the previous equation (\ref{Fequation}).
    
     We may understand the role of the St\"{u}ckelberg fields by reverting to the original vector variables. Then the interaction part of  $\mathcal{L}_{\, 2}'$     in (\ref{stuck2}) reads
\begin{equation}
\label{intl2}
\mathcal{L}_{2I}' \, = \, - e (\mathcal{J}^{\mu} \, + \, \partial^{\mu}\textsl{q})
(A_{\mu} \, + \, \partial_{\mu} \textsl{p}),
\end{equation}
and (\ref{peq}), (\ref{qeq}) have respective counterparts in 
\begin{equation}
\label{peq1}
\partial_{\mu} \mathcal{J}^{\mu} \, + \, \partial^2 \textsl{q}\, = \, 0,
\end{equation}
\begin{equation}
\label{qeq1}
\partial^{\mu} \mathcal{A}_{\mu} \, + \, \partial^2 \textsl{p}\, = \, 0.
\end{equation}
Eliminating  $\textsl{p}$  and $\textsl{q}$  from (\ref{intl2}) with the help of (\ref{peq1}), (\ref{qeq1})
leaves
\begin{equation}
\label{l2eff }
\mathcal{L}_{2I}' \, = \, - e \mathcal{J}^{\mu} \left (\delta_{\mu}^{\nu} \, - \, {\partial_{\mu}\partial^{\nu}
\over \partial^2}\right )  A_{\nu}. 
\end{equation}

This shows that the St\"uckelberg fields ensure that the interaction occurs only between transverse components of   $\mathcal{J}^{\mu}$   and  $A_{\mu}$. 
     For yet another perspective on the role of the St\"{u}ckelberg fields, note that
$-e \int \mathcal{J}^{\mu} A_{\mu}$  
                         is not gauge invariant ( $A_{\mu} \rightarrow A_{\mu} \, + \, \partial_{\mu}\theta$) when 
$\mathcal{J}^{\mu}$   is not conserved. However, the combination   $A_{\mu}  \, + \, \partial_{\mu} \mbox{ \it p}$  is always gauge invariant because  $\mbox{\it p}$ can transform as $\mbox{\it p} \, \mbox{-} \, \theta$. 
      Finally observe that eliminating the St\"{u}ckelberg fields in (\ref{p2equation}) with the help of 
      ({\ref{peq}) and the anomaly equation (\ref{anomaly}) leaves
\begin{equation}
\label{P2eq }
\partial_{\mu} \left ( \mathcal{P}_2 \, + \, {e^2/\pi \over \partial^2} \mathcal{P}_2\right ) \, =\, 0
\end{equation}
This is equivalent to (\ref{p2equation}), but carries no integrability condition.
     Thus we see that the St\"{u}ckelberg modification overcomes difficulties, which arise when the current dual to the axial vector is not conserved.

\section{\mbox{4-Dimensional Model with Topological Mass Generation}}

     For a 4-dimensional generalization of the previous, we adopt the formulation of the 2-dimensional model, presented in Section 2 in terms of the Chern-Pontryagin density and Chern-Simons current, now promoted to four dimensions, 
    $\mathcal{P}_4$ and $\mathcal{C}_4^{\alpha}$   respectively, with the latter coupling to an axial vector current  $\mathcal{J}_{\alpha}^5$      whose divergence is anomalous. The topological entities are constructed from gauge potentials, which we take to be Abelian or non-Abelian;  in either case 
$\mathcal{P}_4$ and $\mathcal{C}_4^{\alpha}$ remain gauge singlets.
\begin{eqnarray}
\label{p4}
\mathcal{P}_4 &\equiv&  {1 \over 2} \epsilon^{\alpha\beta\mu\nu} F_{\alpha\beta}^aF_{\mu\nu}^a \, 
=\, ^*F^{\mu\nu~a} F_{\mu\nu}^a \\ \nonumber
F_{\mu\nu}^a \, &\equiv& \, \partial_{\mu} A_{\nu}^a \, - \partial_{\nu} A_{\mu}^a\, + \, f^{abc} A_{\mu}^bA_{\nu}^c, 
~~~~^*F^{\alpha\beta} \equiv {1\over 2} \epsilon^{\alpha\beta\mu\nu} F_{\mu\nu}
\end{eqnarray}

\begin{equation}
\label{c4}
\mathcal{C}_4^{\alpha} \equiv 2 \epsilon^{\alpha\mu\nu\omega} (A_{\mu}^a\partial_{\nu} A_{\omega}^a
\, + \, {1\over 3} f^{abc} A_{\mu}^aA_{\nu}^bA_{\omega}^c)
\end{equation}
\begin{equation}
\label{c4div}
\partial_{\alpha} \mathcal{C}_4^{\alpha} \, = \, \mathcal{P}_4
\end{equation}
Here   $f^{abc}$  are the structure constants of the appropriate Lie algebra.
    
     Unlike in the 2-dimansional case, the Chern-Simons current is not linear in the gauge vector potential; nevertheless we remain with the potential as the fundamental dynamical variable (see however below). The variation of the Chern-Simons current reads 
\begin{equation}
\label{deltac4}
\delta \mathcal{C}_4^{\alpha} \, = \, 4 ^*F^{\alpha\mu~a}\delta A_{\mu}^a \, - \, 2 \epsilon^{\alpha\nu\omega\mu} \partial_{\nu} (A_{\omega}^a\delta A_{\mu}^a).
\end{equation}
     A further difference from the Schwinger model is that there is no reason to suppose that the dual to the 4-dimensional axial vector current is conserved. On the level of 4-dimensional gamma matrices, the duality relation is
\begin{equation}
\label{epsilon}
\epsilon^{\mu\nu\omega\alpha}\gamma_{\alpha}\gamma^5\, = \, g^{\mu\nu}\gamma^{\omega}\, -\,
g^{\mu\omega}\gamma^{\nu} \, + \, g^{\nu\omega}\gamma^{\mu}\, - \, \gamma^{\mu}\gamma^{\nu} \gamma^{\omega}.
\end{equation}
It is improbable that fermion dynamics (here unspecified) would leave conserved the current dual to the axial vector current. But this is not an obstacle to our construction, because we can employ the St\"{u}ckelberg formalism, as explained in the previous Section, to overcome the difficulty.

     Thus the Lagrange density that we adopt is
\begin{equation}
\label{L4}
\mathcal{L}_4' \, = \, {1\over 2} \mathcal{P}_4^2 \, + \, \Lambda^2 ( \mathcal{C}_4^{\alpha}\, 
+ \, \partial_{\beta} \textsl{p}^{\alpha\beta}) (\mathcal{J}_{\alpha}^5 \, +  
\, \partial^{\gamma} \textsl{q}_{\alpha\gamma}). 
\end{equation}
The St\"{u}ckelberg fields       $\textsl{p}^{\alpha\beta}$    and $\textsl{q}_{\alpha\gamma}$  are anti symmetric in their indices;  $\Lambda^2$     carries mass-squared dimension;  the axial vector current possesses an anomalous divergence.
\begin{equation}
\label{anomaly4}
\partial^{\alpha} \mathcal{J}_{\alpha}^5 \, = \, - N~^*F^{\mu\nu~a}F_{\mu\nu}^a\, = \, -N\mathcal{P}_4
\end{equation}
$N$ is a numerical coupling constant, taken positive.

     Variation of the $\mathcal{L}_4'$  action with respect to $A_{\mu}^a$  gives , with the help of (\ref{deltac4}),
\begin{equation}
\label{varl4 }
\int  \left ( -\partial_{\alpha} \mathcal{P}_4\, + \, \Lambda^2 (\mathcal{J}_{\alpha}^5 \, + \, \partial^{\gamma} \textsl{q}_{\alpha\gamma}) \right ) \delta \mathcal{C}_4^{\alpha} \,  =
\end{equation}
\begin{equation}
\nonumber
\int \left [ 4 \left ( -\partial_{\alpha} \mathcal{P}_4\, + \, \Lambda^2 (\mathcal{J}_{\alpha}^5 \, + \, \partial^{\gamma} \textsl{q}_{\alpha\gamma}) \right )\, ^*F^{\alpha\mu~a}\, - 
2\epsilon^{\alpha\nu\omega\mu} A_{\nu}^a\partial_{\omega}
\left ( -\partial_{\alpha} \mathcal{P}_4\, + \, \Lambda^2 (\mathcal{J}_{\alpha}^5 \, + \, \partial^{\gamma} \textsl{q}_{\alpha\gamma}) \right ) \right ] \delta A_{\mu}^a,
\end{equation}
so that the equation of motion demands 
\begin{equation}
\label{demand}
  2 \left ( -\partial_{\alpha} \mathcal{P}_4\, + \, \Lambda^2 (\mathcal{J}_{\alpha}^5 \, + \, \partial^{\gamma} \textsl{q}_{\alpha\gamma}) \right )\, ^*F^{\alpha\mu~a}\, - 
\epsilon^{\alpha\mu\nu\omega} A_{\nu}^a\partial_{\omega} \, \Lambda^2 (\mathcal{J}_{\alpha}^5 \, + \, \partial^{\gamma} \textsl{q}_{\alpha\gamma}) \, = \, 0.
\end{equation}
Variation of the two St\"uckelberg fields yields the equations
\begin{equation}
\label{deltap}
\partial_{\alpha} (\mathcal{J}_{\beta}^5 \, + \, \partial^{\gamma}\textsl{q}_{\beta\gamma}) \, - \,
\alpha \leftarrow\rightarrow \beta \, =\, 0,
\end{equation}

\begin{equation}
\label{deltaq}
\partial^{\alpha} (\mathcal{C}_4^{\beta} \, + \, \partial_{\gamma}\textsl{p}^{\beta\gamma}) \, - \,
\alpha \leftarrow\rightarrow \beta \, = \, 0.
\end{equation}
The first of these allows setting to zero the second member of (\ref{demand}), while in the first member of that equation we may strip away  $^*F^{\alpha\mu~a}$ with the help of the identity
\begin{equation}
\label{Fidentity}
^*F^{\alpha\mu}F_{\mu\nu} \, = \, -{1\over 4} \delta_{\nu}^{\alpha}\, \mathcal{P}_4.
\end{equation}
Consequently (provided $\mathcal{P}_4 \, \neq 0$) we are left with
\begin{equation}
\label{gradp4}
-\partial_{\alpha} \mathcal{P}_4\, + \, \Lambda^2 (\mathcal{J}_{\alpha}^5 \, + \, \partial^{\gamma} \textsl{q}_{\alpha\gamma})\, = \, 0.
\end{equation}
(Even though we varied $A^a_\mu$, which enters non-lineary into $\mathcal{C}^\alpha_4$, the final equation (\ref{gradp4}) also results by simply varying the composite $\mathcal{C}^\alpha_4$ in (\ref{L4}). This demonstrates the robustness of the derivation.)

The integrability condition on this equation is satisfied by virtue of (\ref{deltap}). Taking another divergence of (\ref{gradp4}) annihilates the St\"uckelberg field because of its anti symmetry, while 
(\ref{anomaly4}) provides the divergence for  $\mathcal{J}_{\alpha}^5$. Thus we are left with 
\begin{equation}
\label{p4final}
\partial^2 \mathcal{P}_4 \, + \, N\, \Lambda^2 \, \mathcal{P}_4 \, = \, 0.
\end{equation}
This shows that the pseudoscalar  $\mathcal{P}_4$  has acquired the mass, $m^{\, 2}  =  N\Lambda^2$. 

By taking the divergence of (\ref{deltap}), we find from (\ref{anomaly4})
\begin{equation}
\label{j5eq}
\mathcal{J}_{\beta}^5 \, + \, \partial^{\gamma} \textsl{q}_{\beta\gamma} \, = \, -{N \over \partial^2} \partial_{\beta}\, \mathcal{P}_4.
\end{equation}
Inserting this in (\ref{gradp4}) yields
\begin{equation}
\label{p4final1}
\partial_{\alpha} \left (  \mathcal{P}_4 \, + \, {N\, \Lambda^2 \over \partial^2} \, \mathcal{P}_4 
\right )  \, = \, 0,
\end{equation}
which  is equivalent to (\ref{p4final}) , but does not entail integrability conditions.

\section{Conclusion}

     While the 4-dimensional transposition of the 2-dimensional Schwinger model succeeds in generating a mass for a pseudoscalar, just as in the 2-dimensional case, there are various shortcomings. To these we now call attention.
     
     The principal defect is the absence of dynamics that should produce the anomaly for the axial vector current. In the Schwinger model, the same dynamics and the same degrees of freedom that generate the mass are also responsible for the anomaly (\ref{anomaly}). In the 4-dimensional theory we must posit the anomaly (\ref{anomaly4}) separately from the mass generating dynamics. Moreover, our final result is that $\mathcal{P}_4$ propagates as a free massive field. Additional dynamics must be specified to describe interactions.
     
     A related question concerns the role in physical theory for our Lagrangian (\ref{L4}). Since it involves dimension eight ($\mathcal{P}_4$) and dimension six ( $\mathcal{C}_4^{\alpha} \mathcal{J}_{\alpha}^5$) operators, it should be viewed as an effective Lagrangian. In this connection, observe that the Born-Infeld action and the radiatively induced Euler-Heisenberg action both contain the Abelian  $(^*F^{\mu\nu}F_{\mu\nu})^2$ quantity in a weak-field expansion [also accompanied by an $(F^{\mu\nu}F_{\mu\nu})^2$ term]. 
     
     The kinetic portion of the Lagrangian in the Weyl ($A_{\ 0} \, = \, 0$ ) gauge involves
$\dot{A}_i \dot{A}_j B^iB^j$  where  $B^i$ is the magnetic field. Canonical analysis and quantization with such a kinetic term faces difficulties because the ``metric'' on  $A_i$ space, {\it viz.}  $B^iB^j$,  is singular. But this poses no problem if our Lagrangian is used for phenomenological purposes, with the semi-classical addition of quantum effects through the chiral anomaly.
   
     The $U(1)$ character of our anomalous current  and the presence in our theory of the Chern-Pontryagin quantity suggest that here we are dealing with the problems of the unwanted axial $U(1)$ symmetry and the mass of the $\eta'$  meson. Conventionally these issues are resolved by instantons \cite{'tHooft:1986nc}. Here we offer a phenomenological description. We relate the axial vector current to the  $\eta\ '$ field,
\begin{equation}
\label{eta}
\mathcal{J}_{\alpha}^5 \, = \, Z \partial_{\alpha} \eta' / \Lambda 
\end{equation}
($Z$ is a normalization) and add an $\eta'$  kinetic term to (\ref{L4}).
\begin{equation}
\label{leta}
\mathcal{L}_{\eta'} \, = \, {1 \over 2} \mathcal{P}_4^2 \, + \, Z \Lambda \mathcal{C}_4^{\alpha}
\partial_{\alpha}\eta' \, + \, {1 \over 2} \partial_{\alpha}\eta'\partial^{\alpha} \eta' 
\end{equation}
[We dispense with the St\"{u}ckelberg fields because the dual of the current in (5.1) 
is conserved.]  Observe that the $\eta'$ field enjoys a constant shift symmetry, as befits the quadratic portion of a Goldstone field Lagrangian. The equations that follow from varying $A_{\ \mu}^{\ a}$   and   $\eta'$  respectively, are       
  \begin{equation}
\label{Aeq}
\partial_{\alpha} (\mathcal{P}_4 \, - \, Z\Lambda \eta') \, ^*F^{\alpha\mu~a} \, = \, 0
\end{equation}
\begin{equation}
\label{etaeq}
\partial^2 \eta' \, + \, Z \Lambda \mathcal{P}_4\, = \, 0
\end{equation}
Together the  two imply
\begin{equation}
\label{p4mass}
\partial^2 \mathcal{P}_4 \, + \, Z^2 \Lambda^2 \mathcal{P}_4\, = \, 0.
\end{equation} 
As before, a mass is generated.

     This may also be seen by rewriting the Lagrangian in (\ref{leta}), apart from a total derivative, as 
\begin{eqnarray}
\label{leta1}
\mathcal{L}_{\eta'} \, &= &\, {1 \over 2} \mathcal{P}_4^2 \, - \, Z \Lambda \mathcal{P}_4\eta' \, + \, {1 \over 2} \partial_{\alpha}\eta'\partial^{\alpha} \eta'\\ \nonumber
 &=& {1 \over 2} (\mathcal{P}_4\, - \, Z\Lambda \eta')^2 \, + \, {1\over 2} \partial_{\alpha}\eta'\partial^{\alpha}\eta' \, - \, {1 \over 2} Z^2\Lambda^2 \eta'^2.
\end{eqnarray}
With  $Z\Lambda \eta'$  absorbed by  $\mathcal{P}_4$,   we see that  $\eta'$ decouples, but carries a mass \cite{Witten:etel}. 

 In the case of 4-dimensional QCD with massless quark flavor(s), equation (\ref{p4mass}) can be obtained 
without any assumptions about the dependence of the effective Lagrangian on the $\eta'$ meson. We only need to assume that the
effective Lagrangian contains the first $\mathcal{P}_{\ 4}^{\ 2}$ term in (\ref{leta1}).  The analog of the second 
term is automatically generated from the anomaly diagram (Fig. 2) that correlates $\mathcal{C}_4^{\alpha}$ and 
$\mathcal{J}_{\alpha}^5$. \\[.5ex]
\centerline{\includegraphics{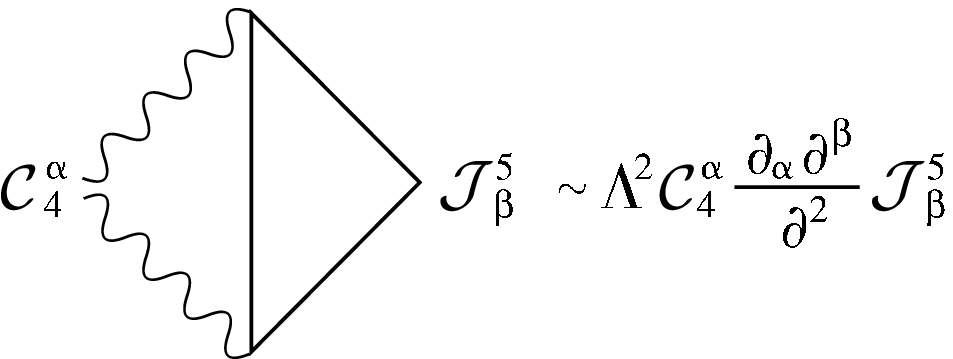}\label{PDFpar}}\\
\centerline{\mbox{\small Figure 2: Anomaly diagram that correlates $\mathcal{C}_4^{\alpha}$} and  $\mathcal{J}_{\beta}^5$}}\\[1.5ex]
 The diagram generates the following operator 
\begin{equation}
\label{contact}
 \Lambda^2 \mathcal{C}_4^{\alpha} {\partial_{\alpha} \partial^{\beta}\over \partial^2} \mathcal{J}_{\beta}^5,
\end{equation}
  where $\Lambda^2$ arises as a momentum cut off.  This expression is also  what one obtains from (\ref{L4}) after eliminating the St\"uckelberg fields 
$\textsl{p}^{\, \alpha\beta}$ and $\textsl{q}_{\alpha\gamma}$   through their equations of motion (4.10), (4.11). 
Thus  massless  quark dynamics due to the anomaly substitute the effect of the  St\"uckelberg fields. 
Variation with respect to $A_{\mu}^a$ yields the analog of  equation (\ref{gradp4}). 
\begin{equation}
\label{gradp4analog}
-\partial_{\alpha} \mathcal{P}_4 \, + \, \Lambda^2 {\partial_{\alpha} \over \partial^2} \partial^{\beta}
\mathcal{J}_{\beta}^5 \, = \, 0
\end{equation}
Using the anomalous divergence relation (\ref{anomaly4}), we arrive to the equation (\ref{p4final1}), 
which is equivalent to (\ref{p4mass}).  Because $\mathcal{P}_4$ acquires a mass, its expectation value 
in the QCD vacuum must vanish. This explains  why 
QCD solves both $U(1)$ and the strong CP problems in the zero quark mass limit \cite{dvali2005}.  

Similar effects should be present in all even dimensions, but the singularity structure and the required dimensional parameter (analog of the 2- and 4-dimensional $e$ and $\Lambda$) will change.

        In conclusion we observe that although both the 2- and 4-dimensional models are formulated in terms of topological entities ($\mathcal{P}, \mathcal{C}^{\, \alpha} $), they are not  topological theories. Examining (3.6), (4.6) we see that the Chern-Simons/axial vector interaction term 
($\mathcal{C}^{\, \alpha} \mathcal{J}_{\alpha}^5$) is a geometric scalar density and can be integrated over a manifold in a diffeomrphism invariant way, without introducing a metric tensor. However, for the  kinetic term ($\mathcal{P}^{\ 2}$) to be a scalar density it must be divided by  $\sqrt{g}$. (In this discussion we ignore the St\"uckelberg terms.) Without this metric factor the theory is not invariant against all diffeomorphisms, but only against the ``volume'' preserving ones with unit Jacobian.

\section*{Acknowledgments}
This work is supported by the U.S.\ Department of Energy 
under cooperative research agreement No.\ DE-FG02-05ER41360.


\vspace{0.5cm}   


\end{document}